\documentclass[10pt,letterpaper]{article}
\usepackage[top=0.85in,left=2.75in,footskip=0.75in]{geometry}

\usepackage{amsmath,amssymb, mathrsfs}

\usepackage{changepage}

\usepackage[utf8x]{inputenc}

\usepackage{textcomp,marvosym}

\usepackage{cite}

\usepackage{nameref,hyperref}

\usepackage[right]{lineno}

\usepackage{microtype}
\DisableLigatures[f]{encoding = *, family = * }

\usepackage[table]{xcolor}

\usepackage{array}

\usepackage{bm}

\usepackage{adjustbox}
\usepackage{array}

\newcolumntype{R}[2]{%
    >{\adjustbox{angle=#1,lap=\width-(#2)}\bgroup}%
    l%
    <{\egroup}%
}

\newcolumntype{+}{!{\vrule width 2pt}}

\newlength\savedwidth

\usepackage{setspace} 
\raggedright
\usepackage[aboveskip=1pt,labelfont=bf,labelsep=period,justification=raggedright,singlelinecheck=off]{caption}

\makeatletter
\renewcommand{\@biblabel}[1]{\quad#1.}
\makeatother

\usepackage{lastpage,fancyhdr,graphicx}
\usepackage{epstopdf}
\fancyheadoffset[L]{2.25in}
\fancyfootoffset[L]{2.25in}
\newcommand{\TW}[1]
{#1}

\newcommand{\ES}[1]
{#1}

\newcommand{\EH}[1]
{#1}

\newcommand{\MB}[1]
{#1}

\usepackage{soul}
\setstcolor{red}

\begin{document}
\vspace*{0.2in}

\begin{flushleft}
{\Large
\textbf\newline{Identifying and characterizing extrapolation in multivariate response data } 
}
\newline
\\
Meridith L Bartley\textsuperscript{1*},
Ephraim M Hanks\textsuperscript{1}, Erin M Schliep\textsuperscript{2}, Patricia A Soranno \textsuperscript{3}, Tyler Wagner \textsuperscript{4}
\\
\bigskip
\textbf{1} Department of Statistics, Pennsylvania State University, University Park, PA, U.S.A.
\\
\textbf{2} Department of Statistics,  University of Missouri, Columbia, MO, U.S.A.
\\
\textbf{3} Department of Fisheries and Wildlife, Michigan State University, East Lansing, Michigan, U.S.A. 
\\
\textbf{4} U.S. Geological Survey, Pennsylvania Cooperative Fish and Wildlife Research Unit, Pennsylvania State University, University Park, Pennsylvania, U.S.A
\\
\bigskip

%
%
These authors contributed equally to this work.





* bartley@psu.edu \MB{(MB)}

\end{flushleft}
\section*{Abstract}

\MB{Faced with limitations in data availability, funding, and time constraints\ES{,} ecologists are often tasked with } making predictions beyond the range of \MB{their} data. In ecological studies, it is not always obvious when and where extrapolation occurs because of the multivariate nature of the data. Previous work on identifying extrapolation has focused on univariate response data, but these methods are not directly applicable to multivariate response data, which are common in ecological investigations. In this paper, we extend previous work that identified extrapolation by applying the predictive variance from the univariate setting to the multivariate case. \MB{We propose using the trace or determinant of the predictive variance matrix to obtain a scalar value measure that\TW{,} when paired with a selected cutoff value\TW{,} allows for delineation between prediction and extrapolation.} We illustrate our approach through an analysis of jointly modeled lake nutrients and indicators of algal biomass and water clarity in over 7000 inland lakes from across the Northeast and Mid-west US. In addition, we \MB{outline} novel exploratory approaches for identifying regions of covariate space where extrapolation is more likely to occur using classification and regression trees. \MB{The use of our Multivariate Predictive Variance (MVPV) measures and multiple cutoff values when exploring the validity of predictions made from multivariate statistical models can help guide ecological inferences.}

Disclaimer: This draft manuscript is distributed solely for purposes of scientific peer review. Its content is deliberative and predecisional, so it must not be disclosed or released by reviewers. Because the manuscript has not yet been approved for publication by the US Geological Survey (USGS), it does not represent any official finding or policy. 



\section*{Introduction}

\begin{figure}[h]
   \centerline{\includegraphics[width=4in]{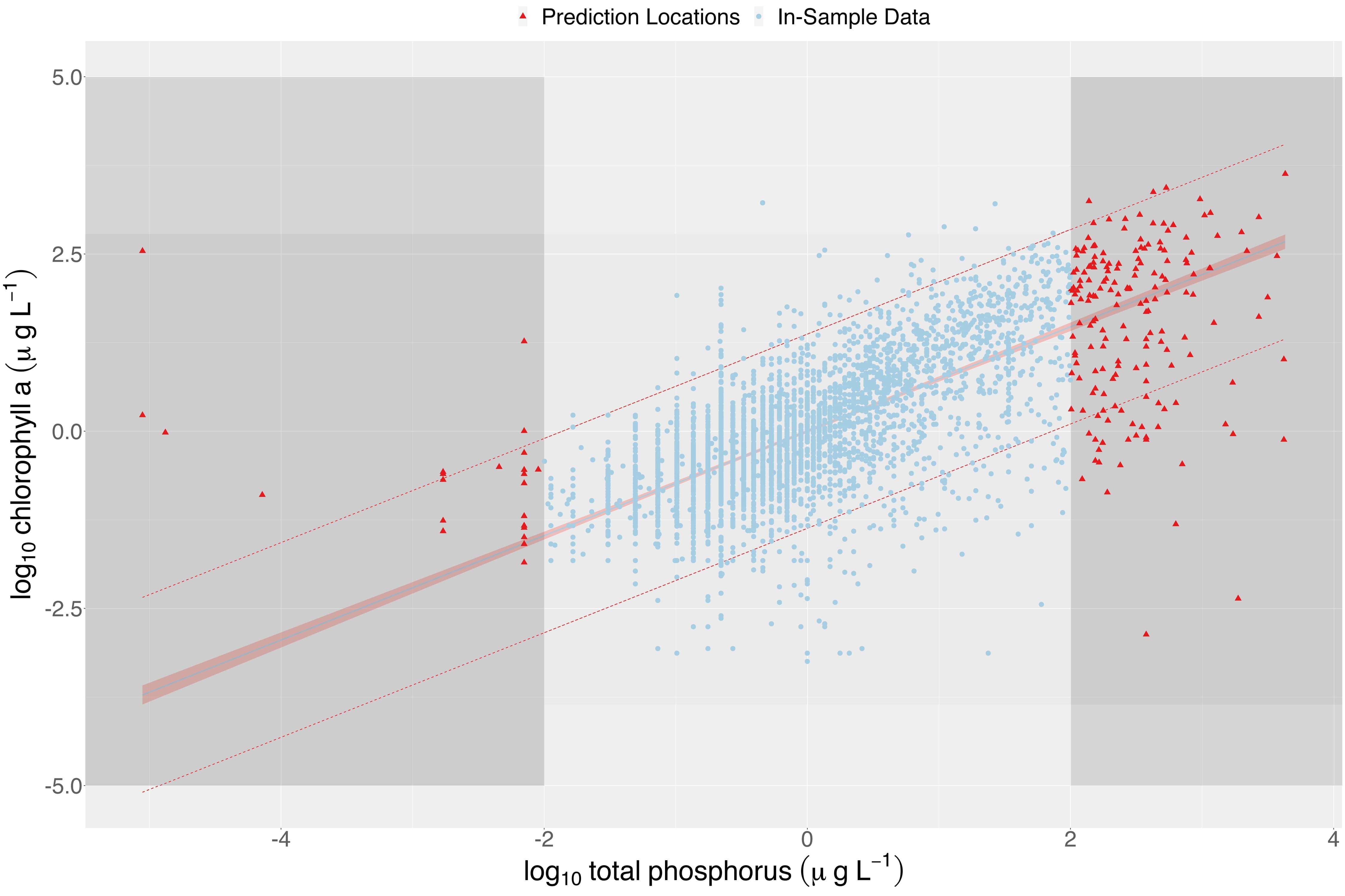}}
\caption{Fit of Chl a–TP relationship for inland lakes using linear regression. A 95\% confidence interval of the mean is included around the regression line. Dashed red lines represents the 95\% prediction interval. Areas shaded in darker grey indicate regions of extrapolation (using the maximum leverage value ($h_{ii}$) to identify the boundaries).}
\label{f:ExtCon}
\end{figure}

\TW{The use of ecological modeling to translate} \MB{observable patterns in nature into quantitative predictions  is vital for scientific understanding, policy making, and ecosystem management. However, generating valid predictions requires robust information across a well-sampled system which is not always feasible given constraints in gathering and accessing data.} Extrapolation is defined as a prediction from a model \TW{that} is \TW{a} projection, extension, or expansion \MB{of} an estimated model (e.g. regression equation, or Bayesian hierarchical model) beyond the range of the data set used to fit that model \cite{Miller2004}. When we use a model fit on available data to predict a value or values at a new location\ES{,} it is important to consider how dissimilar this new observation is to previously observed values. If some or many covariate values of this new point are dissimilar enough from those used when the model was fitted (i.e. either because they are outside the range of individual covariates or because they are a novel combination of covariates) predictions at this point may be unreliable. Fig \ref{f:ExtCon}, adapted from work by Filstrup et al. \cite{Filstrup2014}, illustrates this risk with a simple linear regression between the log transformed measurements of total phosphorous (TP) and chlorophyll \emph{a} (Chl a) in U.S. lakes. The data shown in blue were used to fit a linear model with the estimated regression line shown in the same color. While the selected range of data may be reasonably approximated with a linear model, \MB{t}he linear trend does not extend into more extreme values, and thus our model and predictions are no longer appropriate.   

While ecologists and other scientists know the risks associated with extrapolating beyond the range of their data, they are often tasked \TW{with making} predictions beyond the range of the available data in efforts to understand processes at broad scales, or to make predictions about the effects of different policies or management actions in new locations. Forbes and Carlow \cite{Valery2002a} discuss the double-edged sword of supporting cost-effective progress while exhibiting caution for potential misleading results that would hinder environmental protections. They outline the need for extrapolation to balance these goals in ecological risk assessment. Other works \cite{Freckleton2004a, Colwell11994a, Peters2004a} explore strategies for \MB{addressing the problem of} ecological extrapolation, often in space and time, across applications in management tools and estimation practices. 
Previous work on identifying extrapolation includes Cook's early work on detecting outliers within a simple linear regression setting \cite{Cook1977} and recent extensions to GLMs and similar models by Conn et al. \cite{Conn2015}. The work of Conn et al. defines extrapolation as making predictions that occur outside of a generalized independent variable hull (gIVH), defined by the estimated predictive variance of the mean at observed data points. This definition allows for predictions to be either interpolations (inside the hull) or extrapolations (outside the hull). 

However, the work of Conn et al. \cite{Conn2015} is restricted to univariate response data, which does not allow for the application of these methods to multivariate response models. This is an important limitation because many ecological and environmental research problems are inherently multivariate in nature. Elith and Leathwick \cite{Elith2009} note the need for additional extrapolation assessments of fit in the context of using species distribution models (SDMs) for forecasting across different spatial and temporal scales. Mesgaran et al. \cite{Mesgaran2014} developed a new tool for identifying extrapolation using the Mahalanobis distance to detect and quantify the degree of dissimilarity for points either outside the univariate range or form\MB{ing} novel combinations of covariates. 

In our paper, we present a general framework for quantifying and evaluating extrapolation in multivariate response models that can be applied to a broad class of problems. Our approach may be succinctly summarized as follows: 
\begin{enumerate}
    \item Fit an appropriate model to available multi-response data.
    \item Choose a numeric measure associated with extrapolation that provides a scalar value in a multivariate setting. 
    \item Choose a cutoff or range of cutoffs for extrapolation/interpolation.
    \item Given a cutoff, identify  locations that are extrapolations.
    \item Explore where extrapolations occur. Use this knowledge to help inform future analyses 
    and predictions.
\end{enumerate}
We draw on extensive tools for measures of leverage and influential points to inform decisions of a cutoff between extrapolation and interpolation. We illustrate our framework through an application of this approach on jointly modeled lake nutrients, productivity, and water clarity variables in over 7000 inland lakes from across the Northeast and Mid-west US.

\section*{Predicting \MB{l}ake \MB{n}utrient and \MB{p}roductivity \MB{v}ariables}\label{s:LAGOS}

Inland lake ecosystems are threatened by cultural eutrophication, with excess nutrients such as nitrogen (N) and phosphorus (P) resulting in poor water quality, harmful algal blooms, and negative impacts to higher trophic levels \cite{Carpenter1998}. Inland lakes are also critical components in the global carbon (C) cycle \cite{Tranvik2018}. Understanding the water quality in lakes allows for informed ecosystem management and better predictions of the ecological impacts of environmental change. Water quality measurements are collected regularly by federal, state, local, and tribal governments, as well as citizen-science groups trained to sample water quality. 

The LAGOS-NE database is a multi-scaled geospatial and temporal database for thousands of inland lakes in 17 of the most lake-rich states in the eastern Mid-west and the Northeast of the continental United States \cite{Soranno2017}. This database includes a variety of water quality measurements and variables that describe a lake's ecological context at multiple scales and across multiple dimensions (such as hydrology, geology, land use, and climate).    
%
%

Wagner and Schliep \cite{Wagner2018b} jointly modelled lake nutrient, productivity, and clarity variables and found strong evidence these nutrient-productivity variables are dependent\EH{. They also found that} predictive performance was greatly enhanced by explicitly accounting for the multivariate nature of these data. Filstrup et al. \cite{Filstrup2014} more closely examined the relationship between Chl a and TP and found nonlinear models fit the data better than a log-linear model. Most notably for this work, the relationship of these variables differ in the extreme values of the observed ranges; while a linear model may work for a moderate range of these data it is imperative that caution is shown before extending results to more extreme values (i.e., to extremely nutrient-poor or nutrient-rich lakes). 

\begin{figure}[h]
   \centerline{\includegraphics[width=5in]{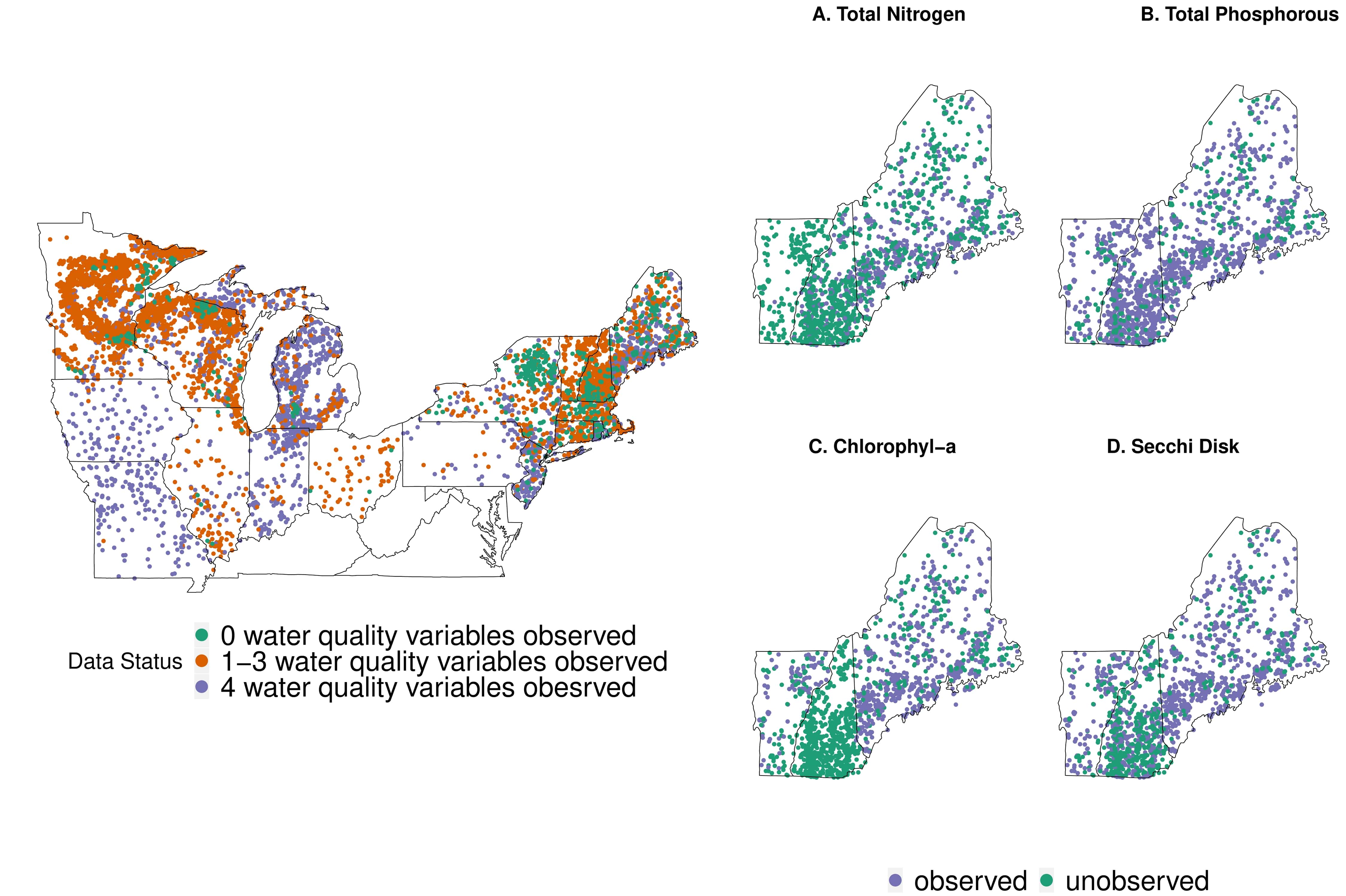}}
\caption{Left: map of inland lake locations with full, partial, or missing response variables. Missing response variables are lakes where all water quality measures have not been observed, while partial status indicates only some lake response variables are unobserved. Covariates were quantified for all locations. Right: subset of data status (observed or missing) for each response variable. \MB{All spatial plots in this paper were created using the Maps package \cite{Becker2013Maps:Maps} in R to provide outline of US states.}} 
\label{f:missing}
\end{figure}

In this study, following Wagner and Schliep, we consider four variables: total phosphorous (TP), total nitrogen (TN), \TW{Chl a}, and Secchi disk depth (Secchi) as joint response variables of interests. Each lake may have observations for all four of these variables, or only a subset. 
Fig \ref{f:missing} shows response variable availability (fully observed, partially observed, or missing) for each lake in the data set. A partially observed set of response variables for a lake indicates that at least one, but not all, of the water quality measures were sampled. We consider several covariates at the individual \TW{lake} and watershed scales as explanatory variables including maximum depth (m), mean base flow (\%), mean runoff (mm/yr), road density (km/ha), elevation (m), stream density (km/ha), the ratio of watershed area to lake area, and the proportion of forested and agricultural land in each lake's watershed. One goal among many for developing this joint model is to be able to predict TN concentrations for all lakes across this region, and eventually the entire continental US. Our objective is to identify and characterize when predictions of these multivariate lake variables are extrapolations. To this end, we will review and develop methods for identifying and characterizing extrapolation in multivariate settings.   
 
\section*{Materials and \MB{m}ethods} 

\subsection*{Review of \MB{c}urrent \MB{w}ork}
\subsubsection*{Cook's \MB{i}ndependent \MB{v}ariable \MB{h}ull}
As this work builds upon the work of Cook \cite{Cook1977} and Conn et al. \cite{Conn2015}, we start with a review of their independent variable hull (IVH) and generalized independent variable hull (gIVH) approaches. Cook's work focuses on the identification of influential points in a linear regression setting. A linear regression model is written as
\begin{linenomath}
\begin{equation}
	\mathbf{y} = \mathbf{X}\bm{\beta} + \bm{\epsilon}
\end{equation}
\end{linenomath}
where $\mathbf{y} = [y_1, \dots, y_n]'$ denotes a vector of $n$ univariate observed responses, $\mathbf{X}$ denotes the covariate matrix with an intercept, $\bm{\beta}$ are the covariate coeffecients, and $\bm{\epsilon}$ are independent, mean-zero normally distributed residuals. \MB{Throughout this paper, we use bold lowercase letters to denote vectors and bold uppercase letter to denote} \TW{matrices}. The predicted value of $\mathbf{y}$ may be calculated 
\begin{linenomath}
\begin{equation}
    \hat{\mathbf{y}} = \mathbf{X} \hat{\bm{\beta}}
\end{equation}
\end{linenomath}
where $\hat{\bm{\beta}}$ may be replaced with its OLS estimate ($\hat{\bm{\beta}} = (\mathbf{X}'\mathbf{X})^{-1}\mathbf{X}'\mathbf{y}$) to obtain
\begin{linenomath}
\begin{equation}
    \hat{\mathbf{y}} = \mathbf{X}(\mathbf{X}'\mathbf{X})^{-1}\mathbf{X}' \mathbf{y}
\end{equation}
\end{linenomath}
\ES{Equivalently, the} hat matrix, $\mathbf{H} = \mathbf{X(X'X)}^{−1} \mathbf{X}'$,  when multiplied by the observed $\mathbf{y}$ vector will produce the predicted values. \MB{T}he predicted response for observation $i$ can be written as a linear combination of the $n$ response variables,
\begin{linenomath}
\begin{equation}
    \hat{y}_{i}=h_{i 1} y_{1}+h_{i 2} y_{2}+\ldots+h_{i i} y_{i}+\ldots+h_{i n} y_{n} \quad \text { for } i=1, \ldots, n.
\end{equation}
\end{linenomath}
The diagonal elements of the hat matrix ($h_{ii} = \mathbf{x}_i′(\mathbf{X}′\mathbf{X})^{−1}\mathbf{x}_i$) are called leverages, and while they only depend on the explanatory variables, they indicate the influence observations, $y_i$, have on their own predicted values, $\hat{y}_i$.  A higher leverage $h_{ii}$ indicates a higher influence of ${y}_i$ in determining the model fitted response $\hat{{y}}_i$. This relationship means leverage values are useful quantities to explore when looking for \EH{influential points}. The corresponding residual vector is $\mathbf{r} = \mathbf{y} - \hat{\mathbf{y}} = (\mathbf{I} - \mathbf{H})\mathbf{y}$. Building on confidence ellipsoids for multiple coefficients, Cook's Distance, $D_i$, is a measure to explore the individual contribution of the $i$\textsuperscript{th} data point in a linear regression analysis.
This measure may be calculated by
\begin{linenomath}
\begin{equation}
 \begin{aligned}
    D_i &= \frac{(\hat{\bm{\beta}}_{-(i)} - \hat{\bm{\beta}})' \mathbf{X}'\mathbf{X} (\hat{\bm{\beta}}_{-(i)} - \hat{\bm{\beta}}) }{p s^2} \\
     &= \frac{t_i^2}{p} \left(\frac{h_{ii}}{1 - h_{ii}}\right)
 \end{aligned}
\end{equation}
 \end{linenomath}
where $p$ represents the number of parameters, $s^2$ is $\mathbf{r}'\mathbf{r}/(n - p)$ , and $t_i$ is the $i$\textsuperscript{th} studentized residual. We use $\hat{\bm{\beta}}_{-(i)}$ to indicate the estimate of the the ${\bm{\beta}}$ vector without the $i$\textsuperscript{th} data point.  With all other values held constant, this measure increases as a function of the ratio of $h_{ii}$ over $1 - h_{ii}$, which depends only on the design points within $\mathbf{X}$. As such, Cook defines his independent variable hull (IVH) as the smallest convex set containing all of the design points. Let $h$ denote the maximum diagonal element of this hat matrix (i.e., $h = \max(\text{diag}(\mathbf{H})))$, then a new observation, $\mathbf{x}_0$, is within this defined IVH whenever
\begin{linenomath}
\begin{equation}
	\mathbf{x}_0'\mathbf{(X'X)^{-1}x}_0 \leq h
\end{equation}
\end{linenomath}
and predicting at a point beyond the hull will imply an extrapolation.

The hat matrix and its diagonals are useful diagnostics for finding outliers in a linear regression setting. Similarly, \EH{the} Mahalanobis distance (MD) \cite{Mahalanobis1936} can be used for identifying outliers. MD and leverage are monotonically related, as the scale-invariant  squared MD may be represented by  
\begin{linenomath}
\begin{equation}
 \begin{aligned}
 MD^2_i &= (\mathbf{x}_\MB{i} - \bar{\mathbf{x}})\hat{\bm{\Sigma}}^{-1}(\mathbf{x}_\MB{i} - \bar{\mathbf{x}})'\\
 &= (\MB{l} - 1) (h_{ii} - \frac{1}{\MB{l}})    
 \end{aligned}
\end{equation}
\end{linenomath}
where $\mathbf{x}_\MB{i}$ is a data point (with $p$ total covariate observations), $\bar{\mathbf{x}}$ is the mean vector for all $\mathbf{x}$ (i.e. $\bar{x}_i = \frac{1}{\MB{l}}\sum^\MB{l}_{i = 1} x_i$ where \MB{$l$} is the number of observed lakes
), and $\hat{\bm{\Sigma}}$ is the sample covariance matrix. We assume $\bar{\mathbf{x}} = 0 $ without loss of generality. This relationship assumes the model matrix $\mathbf{X}$ includes an intercept and makes use of the following \EH{partitioning}
\begin{linenomath}
\begin{equation}
    \mathbf{(X^\prime X)}^{-1} = 
    \begin{pmatrix}
    \frac{1}{\MB{l}} & \mathbf{0}^\prime \\ 
    \mathbf{0} & \frac{1}{\MB{l}-1}\hat{\bm{\Sigma}}^{-1}
    \end{pmatrix}.
\end{equation}
\end{linenomath}
\MB{We may work backwards from $h_{ii}$ to obtain
\begin{linenomath}
\begin{equation}
 \begin{aligned}
h_{ii} &= (1; \mathbf{x}_i)  \begin{pmatrix}
    \frac{1}{l} & \mathbf{0}^\prime \\ 
    \mathbf{0} & \frac{1}{l-1}\hat{\bm{\Sigma}}^{-1}
    \end{pmatrix}
(1; \mathbf{x}_i)^\prime \\
&=\frac{1}{l} + \frac{1}{l-1}\mathbf{x}_i \hat{\bm{\Sigma}}^{-1}\mathbf{x}_i^\prime \\
&=\frac{1}{l} + \frac{1}{l-1} MD^2_i.
 \end{aligned}
\end{equation}
\end{linenomath}
}
This definition remains useful without any underlying distributional assumption of the data. For example, empirically obtained quantile cutoff values can serve reasonably well as threshold for declaring outliers. However, for multivariate-normal data, the squared MD \MB{can be transformed into probabilities using a chi-squared cumulative probability distribution \cite{Etherington2019MahalanobisError} such that points that have a very high probability of not belonging to the distribution could be classified as outliers.} In either scenario, outliers can be detected using only predictor variables by calculating $\mathbf{x}_0(\mathbf{X'X})^{-1}\mathbf{x}_0$ and comparing with $\max(\text{diag}(\mathbf{X}(\mathbf{X'X})^{-1}\mathbf{X}))$.



\subsubsection*{Conn's \MB{g}eneralized IVH}
The work of Cook does not immediately extend to generalized linear models (GLMs) where the assumption of Gaussian errors is relaxed. 
To extend to the GLM case, Conn et al. \EH{(2015)} define a generalized independent variable hull (gIVH) for a generalized linear model,
\begin{linenomath}
\begin{equation}
    \MB{y}_i \sim f_\MB{y} (g^{-1}(\mu_i)),
\end{equation}
\end{linenomath}
where $f_\MB{y}$ denotes a probability density or mass function, $g$ gives the necessary link function, and $\mu_i$ is a linear predictor (e.g. $\mu_i = \mathbf{x}_i'\bm{\beta}$). Using Cook's IVH boundary connection to predictive variance, Conn et al. define the gIVH as the set of all predicted locations $\mathscr{L}_P$ for which
\begin{linenomath}
\begin{equation} \label{eq:gIVH}
	\text{var}(\hat{y}_{\MB{p}} | \mathbf{y}) \leq \max_{\MB{o} \in \mathscr{L}_O}[\text{var}(\hat{y}_{\MB{o}}|\mathbf{y})]
\end{equation}
\end{linenomath}
where $\MB{p} \in \mathscr{L}_P$, $\hat{y}_\MB{p} = g^{-1}(\mathbf{x}_\MB{p}\mathbf{\hat{\beta}})$ corresponds to the mean prediction at $\MB{p}$, $\mathscr{L}_O$ denotes the set of locations where data are observed, and ${\hat{y}}_\MB{o}$ denotes predictions of observations at $\mathbf{x}_\MB{o} \in \mathscr{L}_O$. \MB{In addition to this approach of using $o$ and $p$ to index observed and predicted locations, respectively, we will also in this paper use $i$ to index the collective set of locations (i.e. $i \in \mathscr{L}_o \cup \mathscr{L}_p$).}
The variance of this predictive mean when a non-identity link is used may be found using the delta method which may be written as
\begin{linenomath}
\begin{equation}
 \begin{aligned}
    \text{var}({\hat{y}_i}) &= \text{var}(g({\hat{\mu}_i}))\\
    &\approx \bm{\Delta} \text{var}({\hat{\mu}_i}) \bm{\Delta}',
\end{aligned}
\end{equation}
 \end{linenomath}
where $\bm{\Delta}$ is a matrix of partial derivatives of the function $g(\mu)$ with respect to its parameters, evaluated at the estimators, $\hat{\mu}$.

\subsubsection*{Prediction \MB{v}ariance}

The IVH approach of Cook's work uses only the design matrix, $\mathbf{X}$, to calculate the hat matrix, $\mathbf{H}$. Since the hat matrix is not always well defined for more complicated models, prediction variance may be substituted as a boundary for Conn et al.'s gIVH.
This Prediction Variance (PV) approach requires the design matrix, $\mathbf{X}$, in addition to the response variable \MB{vector, $\mathbf{y}$}. Finding the prediction variance under a univariate response model is accomplished by either direct calculation of $\text{var}(\bm{\hat{y}})$ or through posterior predictive inference resulting in a single scalar value for each location. 

Writing our linear predictor generally as 
\begin{linenomath}
\begin{equation}
    \bm{\mu} = \mathbf{X}\bm{\beta}
\end{equation}
\end{linenomath}
where $\mathbf{X}$ is the design matrix and $\bm{\beta}$ is the vector of unknown parameters to be estimated, we find 
\begin{linenomath}
\begin{equation}
    \text{var}(\bm{\hat{\mu}}) = \mathbf{X} \text{var}(\bm{\hat{\beta}}) \mathbf{X'}.
\end{equation}
\end{linenomath}
Under a linear model\MB{,}
\begin{linenomath}
\begin{equation}
    \bm{\hat{\beta}} = (\mathbf{X}'\mathbf{X})^{-1}\mathbf{X}'\mathbf{y}
\end{equation}
\end{linenomath}
where the distribution of $\bm{\hat{\beta}}$ is 
\begin{linenomath}
\begin{equation}
    \bm{\hat{\beta}} \sim \text{N}(\bm{\beta}, \sigma^2 (\mathbf{X}'\mathbf{X})^{-1})
\end{equation}
\end{linenomath}
and thus $\text{var}(\bm{\hat{\mu}}) = \sigma^2 \mathbf{X} (\mathbf{X}'\mathbf{X})^{-1}\mathbf{X}' $ is proportional to the hat matrix used in Cook's IVH criteria. 

\MB{While this extrapolation approach can be applied under inference in both frequentist and Bayesian approaches, we focus on} a Bayesian setting \MB{in which} we may calculate the prediction variance using the posterior predictive distribution of a\MB{n out-of-sample} observation, ${y}_{\MB{p}}$, given the observed data, $\mathbf{y}$. Using $[\cdot]$ to denote a probability distribution, this distribution is
\begin{linenomath}
\begin{equation}
    [{y}_{\MB{p}} | \mathbf{y}] = \int [{y}_{\MB{p}} | \bm{\theta}][\bm{\theta} | \mathbf{y}] d\bm{\theta}
\end{equation}
\end{linenomath}
where $\bm{\theta} = (\bm{\beta}, \sigma^2)$ in our linear model and $[\bm{\theta} | \mathbf{y}]$ is the posterior distribution.  
We may approximate the posterior predictive distribution through MCMC by sampling ${y}_{\MB{p}}^{(a)} \sim [{y}_\MB{p} | \bm{\theta}^{(a)}]$ using $\bm{\theta}^{(a)}$ at each iteration ($a = 1, \dots, A)$ of the algorithm. With the posterior predictive distribution and with  observed covariates, ${\mathbf{x}_\MB{p}}$ at each new location we may calculate $  {\mu}^{(a)} = \mathbf{x}_\MB{p}'\bm{\beta}^{(a)}$  at each MCMC iteration 
and Monte Carlo predictive inference can be obtained using $\hat{{y}}_\MB{p}^{(a)} = {\mu}^{(a)}$ for the converged MCMC samples. The prediction variance may be approximated by \begin{linenomath}
\begin{equation}
\label{e:scalarPV}
    \hat{\text{var}}(\hat{{y}}_\MB{p} | \mathbf{y}) = \frac{\sum^A_{a = 1} (\hat{{y}}_\MB{p}^{(a)} - \text{E}(\hat{{y}}_\MB{p} | \mathbf{y}))^2}{A}.
\end{equation}
\end{linenomath}
With this sample-based calculation of prediction variance for our measure of extrapolation we can easily extend this univariate approach to the multivariate setting.

\subsection*{Extension to the \MB{m}ultivariate \MB{c}ase}\label{s:MVcase}

Building upon this previous work, we aim to extend measures of extrapolation to handle predictions of multivariate data. We illustrate this using the inland lake nutrient and productivity data. Following the multivariate linear model developed by Wagner and Schliep \cite{Wagner2018},
the joint nutrient-productivity model can be collectively written as:
\begin{linenomath}
\begin{equation}
    \mathbf{y}_{i}  = \mathbf{B}\mathbf{x}_{i} + \bm{\epsilon}_{i}, \quad \bm{\epsilon}_{i} \stackrel{\text{iid}}{\sim} \text{N}(\mathbf{0}, \bm{\Sigma})
\end{equation}
\end{linenomath}
where $\mathbf{y}_{i}$ denotes a vector column of a matrix, $\mathbf{Y}$,where \MB{$y_{ni}$} is the value of the n\textsuperscript{th} lake nutrient-productivity variable for lake $i$. 
For each lake $i$\MB{,}
 \begin{linenomath}
\begin{equation}
 \begin{aligned}
         \mathbf{y}_{i} &= [\text{TN}_i, \text{TP}_i, \text{CHL}_i, \text{Secchi}_i]'\\
         \mathbf{y}_{i} &= \mathbf{B}\mathbf{x}_{i} + \bm{\epsilon}_{i} \Leftrightarrow {y}_{\MB{ni}} = \mathbf{\MB{b}}'_{n }\mathbf{x}_{i} + \epsilon_{\MB{ni}}
 \end{aligned}
\end{equation}
 \end{linenomath}
where $\mathbf{B}$ is a matrix of coefficients such that \MB{$\mathbf{b}'_{n}$ is a row vector where} ${\MB{b}}_{nq}$ is the coefficient of the $\MB{q}$\textsuperscript{th} predictor variable for the $n$\textsuperscript{th} lake nutrient response variable. \MB{The notation $\mathbf{x}_i$ represents a $q \times 1$ vector of predictor variables for lake $i$.} 
Here, again for lake $i$,  $\bm{\epsilon}_i \sim \text{N}(\bm{0}, \bm{\Sigma})$ where $\bm{\Sigma}$ is a $n$ x $n$ covariance matrix capturing the dependence between nutrient-productivity variables not accounted for by the regression. We assume multivariate errors are independent and identically distributed across lakes. Following Wagner and Schliep (2018), we take a Bayesian approach and specify priors for all model parameters. 
\begin{linenomath}
\begin{equation}
\begin{aligned}
    \MB{b_{rc}} &\stackrel{iid}{\sim} \text{N}(0, 100), \quad \MB{r} = 1, \dots, n, \MB{c} = 1, \dots, \MB{q}\\
       \bm{\Sigma} &\sim \text{IW(I, \MB{q}+1)}
\end{aligned}
\end{equation}
 \end{linenomath}
Prediction\EH{s} \MB{of different response types
covary in multivariate models}, complicating our definition of a gIVH (see Eq \ref{eq:gIVH}) which relies on finding a maximum univariate value. \EH{Where a univariate model would yield a scalar prediction variance (Eq \ref{e:scalarPV}), a multivariate model will have a prediction covariance matrix.} We propose capturing the size of a covariance matrix using univariate measures. Note this is similar to A-optimality and D-optimality criteria used in experimental design \cite{Gentle2007}.  

Further, using our novel numeric measure of extrapolation, we aim to take advantage of the multivariate response variable information to explore when we may identify an additional observation's (i.e. covariates for a new lake location) predictions as extrapolations for all response values\EH{, jointly}. We also \EH{present an approach to} identify when we cannot trust a prediction for only a single response variable at either a new lake location, or a currently partially-sampled lake.
The latter identification would be useful for a range of applications in ecology. For example, in the inland lakes project, one important goal is to predict TN because this essential nutrient is not well-sampled across the study extent, and yet is important for understanding nutrient dynamics and for informing eutrophication management strategies for inland lakes. In this case, to accommodate TN \MB{not being} observed (i.e. sampled) as often as some other water quality variables, we can leverage the knowledge gained from samples of other water quality measures taken more often than TN (e.g. Secchi disk depth \cite{Lottig2014}  is a common measure of water clarity obtained on site, while other water quality measurements require samples to be sent \TW{to a lab} for analysis). We first outline our approach for identifying extrapolated new observations using a measure of predictive variance for lakes that have been fully or partially sampled and used to fit a model. Then, we describe how this approach can be applied to the prediction of TN in lakes for which it has not been sampled.

\subsubsection*{Multivariate \MB{e}xtrapolation \MB{m}easures} \label{ss:MVPV}
Using available data for both complete and partial measurements of water quality at inland lake observations (\MB{here,} $\mathbf{Y}  = \{\mathbf{y}_\MB{o}\MB{, o} \in \mathscr{L}_O\}$) and corresponding covariates of these sampled locations ($\mathbf{{X}}$) we first fit an appropriate model to obtain estimates for parameters needed for prediction (here, $\hat{{\mathbf{B}}}$ and $\hat{\bm{\Sigma}})$. 
With these values\MB{, in addition to covariates that correspond with unsampled locations,} we may either directly calculate the prediction variance or, in a Bayesian setting, simulate it via posterior predictive inference.  We denote this prediction variance with $\mathbf{V}_i$ where
\begin{linenomath}
\begin{equation}
    \begin{aligned}
    \mathbf{V}_i &= \text{var}(\hat{{\bm{y}}}_{i} |\mathbf{Y})\\
    &=\text{var}(\mathbf{Bx}_{i} | \mathbf{Y} ).
\end{aligned}
\end{equation}
 \end{linenomath}
Each $\mathbf{V}_i$ is a square matrix for a sampled or unobserved location, (i.e. the combined sets of $\mathscr{L}_O$ and $\mathscr{L}_P$, respectively), with the dimensions equal to the number of response variables in the model. As in the univariate case, we propose to characterize extrapolation by comparing prediction variances of unobserved lakes with corresponding prediction variances of observed lakes. To obtain a scalar value representation of each covariance matrix we propose using the trace or determinant. In this paper, we will refer to these multivariate posterior variance (MVPV) measures for each inland lake observation with respect to how this scalar value representation is calculated:
\begin{linenomath}
\begin{equation}
 \begin{aligned}
     \text{MVPV(tr)\textsubscript{i}} &= tr(\mathbf{V}_i) = \sum^{4}_{ n = 1} \MB{\mathbf{V}}_i[n, n] \text{ where } \MB{\mathbf{V}}_i[n, n] = n \text{\textsuperscript{th} diagonal element of } \MB{\mathbf{V}}_i\\ \text{MVPV(D)\textsubscript{i}} &= |\mathbf{V}_i|= \prod_{n = 1}^4 \lambda_{in} \text{ where } \lambda_{in} = n\text{\textsuperscript{th} Eigenvalue of } \mathbf{V}_i
     \label{e:vi}
\end{aligned}
\end{equation}
 \end{linenomath} 
The trace (tr) of an $n\times n$ square matrix $\MB{\mathbf{V}}$ is defined to be the sum of the elements on the main diagonal (the diagonal from the upper left to the lower right). This does not take into account the correlation between variables \MB{and is not a scale-invariant measure. As the response  variables for the inland lakes example are log transformed, we chose to explore the use of this measure for obtaining a scalar value extrapolation measure.} The determinant \TW{(D)} takes into account the correlations among pairs of variables \MB{and is scale-invariant}. In this paper, we explore both approaches by quantifying extrapolation using our multivariate model of the LAGOS-NE lake data set by:
 \begin{enumerate}
   \item Finding the \ES{joint} posterior distribution $\mathbf{B}, \bm{\Sigma} | \mathbf{Y} $
   \item Calculating \ES{the} posterior predictive variance at in-sample lakes
   \item Calculating \ES{the} posterior predictive variance at out of sample lakes
   \item Identifying extrapolations by comparing out of sample MVPV values to a cutoff value chosen using the in-sample values. 
 \end{enumerate}
\subsubsection*{Conditional \MB{s}ingle \MB{v}ariable \MB{e}xtrapolation \MB{m}easures}
\MB{The chosen} numeric measure \MB{of MV} extrapolation includes information from the entire set of responses. In the inland lake example, this could be used to identify unsampled lakes where prediction of the whole vector of response variables (TN, TP, Chl a, Secchi) are extrapolations. However, even when a joint model is appropriate, there are important scientific questions that can be answered with prediction of a single variable. 

To focus on a single response variable (taken to be the $n$\textsuperscript{th} variable without loss of generality) conditioned on others, we now define the conditional multivariate predictive variance (CMVPV) as
\begin{equation}
    \text{CMVPV\textsubscript{i}} = \text{var}(\hat{y}_{\MB{ni}}|\mathbf{y}_{\MB{-ni}}, \mathbf{Y})
\end{equation}
where $\mathbf{y}_{\MB{-ni}}$ are the response variables for the $i$\textsuperscript{th} lake observation being conditioned upon. With the Bayesian approach detailed above, we can get sample realizations of the conditional MVN distribution of $[y_{\MB{ni}}|\MB{\mathbf{y}_{-ni}, }\mathbf{Y}]$ for all MCMC iterations\MB{, however, we still need a way to obtain univariate prediction variance.} \MB{While one could directly sample from $[\hat{y}_{ni}|\mathbf{y}_{-ni}, \mathbf{Y}]$, w}e suggest the following relationship
\begin{linenomath}
    \begin{equation}
        \begin{aligned}
    \text{var}(y_{\MB{ni}}|\mathbf{Y}) &= \text{var}(\hat{y}_{\MB{ni}} + \hat{\epsilon}_{\MB{ni}} |\mathbf{Y})\\
    &= \text{var}(\hat{y}_{\MB{ni}}|\mathbf{Y}) + \text{var}(\hat{\epsilon}_{\MB{ni}} |\mathbf{Y})\\
    &= \text{var}(\hat{y}_{\MB{ni}}|\mathbf{Y}) + \hat{\sigma}_n 
\end{aligned}
\end{equation}
 \end{linenomath} 
Because in our model the multivariate errors $\bm{\epsilon}_\MB{i}$ are independently and identically distributed across lakes, \MB{then} $\text{var}(\hat{\epsilon}_{\MB{ni}}) = \sigma_n$.  As $\sigma_n$ is constant across all lakes, we can use either $\text{var}(y_{\MB{ni}}|\mathbf{Y})$ or $\text{var}(\hat{y}_{\MB{ni}}|\mathbf{Y})$  to characterize extrapolation\MB{. While the variances are different, the conclusions about extrapolation will be the same as both observed and unobserved lakes will have the same constant added}.

\MB{As the inland} lake data \MB{are modelled with a} multivariate normal (MVN) \MB{distribution}, we may use results from a conditional MVN distribution\MB{.} \MB{I}f ${\mathbf{y}}_{i}$ is jointly normally distributed as 
\begin{equation}
\mathbf{y}_{i} \sim N(\bm{\mu}_{i}, \bm{\Sigma})      
\end{equation}
where $\bm{\mu}_{i} = \mathbf{Bx}_{i}$ \MB{and if we condition the response for one nutrient measure for lake $i$ on all other available nutrient measures for that lake } then, 
 \begin{equation}
 [y_{\MB{ni}}|\mathbf{y}_{\MB{-ni}} = \mathbf{a}, \mathbf{Y}] \sim N(\bar{\mu}, \bar{\Sigma})   
 \end{equation}
 \MB{For a lake observation that has been fully sampled for all four measures, we may compartmentalize the covariance matrix ${\bm{\Sigma}}$ for use in calculating the scalar values $\bar{\mu}$ and $\bar{\Sigma}$ for $[y_{1i} | \mathbf{y}_{(2, 3, 4)i}]$ in the following way, 
\begin{equation}
\bm{\Sigma} = 
\left[
\begin{array}{c|ccc}
\Sigma_{11} & \Sigma_{12} & \Sigma_{13} & \Sigma_{14} \\
\hline 
\Sigma_{21} & \Sigma_{22}& \Sigma_{23}& \Sigma_{24} \\ 
\Sigma_{31} & \Sigma_{32}& \Sigma_{33}& \Sigma_{34} \\ 
\Sigma_{41} & \Sigma_{42}& \Sigma_{43}& \Sigma_{44}
\end{array}
\right]
\equiv \begin{bmatrix}
\Sigma_{11} & \tilde{\bm{\Sigma}}_{12}\\
\tilde{\bm{\Sigma}}_{21} & \tilde{\bm{\Sigma}}_{22}
\end{bmatrix}
\end{equation}
\EH{Within this new configuration of $\bm{\Sigma}$, $\Sigma_{11}$ does not change, however $\tilde{\boldsymbol\Sigma}_{12}$ is the submatrix of $\boldsymbol\Sigma$ of dimension 1$\times$3 containing $\Sigma_{12}, \Sigma_{13},$ and $\Sigma_{14}$, $\tilde{\boldsymbol\Sigma}_{21}$ is the submatrix of dimension 3$\times$1 containing $\Sigma_{21}, \Sigma_{31},$ and $\Sigma_{41}$, and $\tilde{\boldsymbol\Sigma}_{22}$ is the submatrix of dimension 3$\times$3 containing the remaining elements of $\bm{\Sigma}$.}}
\MB{Using this partitioned $\bm{\Sigma}$ we may obtain $\bar{\mu}$ and $\bar{\Sigma}$.
\begin{linenomath}
\begin{equation}
 \begin{aligned}
     \bar{\mu} &= \mu_{ni} - \tilde{\bm{\Sigma}}_{12}\tilde{\bm{\Sigma}}_{22}^{-1}(\mathbf{a} - \bm{\mu}_{-ni})\\
     \bar{\Sigma} &= \Sigma_{11} - \tilde{\bm{\Sigma}}_{12}\tilde{\bm{\Sigma}}_{22}^{-1}\tilde{\bm{\Sigma}}_{21} \label{e:sigmabar}
\end{aligned}
\end{equation}
 \end{linenomath}
 Any of the four response variables may be considered to be variable 1 and so this general partition approach may be used for any variable conditioned on all others.}
The values of $\bm{\mu}_{\MB{-ni}}$ and $\bm{\Sigma}$ are determined by the availability of data for the three variables we are conditioning on. These water quality measure can be fully, partially, or not observed. 

In the instances where all other measures  have not been observed then we may still proceed to calculate $\text{var}(\hat{y}_{\MB{ni}}|\mathbf{Y})$ as done for the MVPV. In order for this measure of variance to be comparable to other CMVPV values, we must add $\text{var}(\hat{\epsilon}_{\MB{ni}}) = \sigma_n$. This Conditional MVPV (CMVPV) measure results in a single scalar value for each location, $i \in \mathscr{L}_O \cup \mathscr{L}_P$, that may be used as outlined above to diagnose extrapolation. 

\subsection*{Cutoffs vs \MB{c}ontinuous \MB{m}easures}

With our selection of multivariate prediction variance measures (MVPV(tr), MVPV(D), and CMVPV) we may proceed by choosing a cutoff or range of cutoffs for delineating between extrapolation and interpretation. The role of a cutoff value or criteria in identifying and characterizing extrapolation is to delineate between prediction and extrapolation. Or rather, where (among covariate values, time, space, etc) may we expect our model to provide accurate prediction values versus where should we exhibit caution when \MB{using} model-based predictions. A key decision for whether or not we label a prediction as an extrapolation (and thus identifying the location as a potentially unreliable extension of our model beyond the data) is the measure used as a boundary cutoff. Previous work \cite{Cook1979, Conn2015} has used the maximum prediction variance as the cutoff of the g(IVH). However, many datasets contain outliers and influential points -- data locations very different from the rest of the data. Choosing a cutoff for extrapolation based on the most extreme outlier in a data set will result in a very conservative definition of extrapolation for many datasets. We thus \TW{suggest} (and illustrate below) a range of extrapolation cutoffs be explored, resulting in a more complete understanding of potential extrapolation. Each cutoff value we propose is a function of the scalar value\EH{d} prediction variance representations of MVPV(D or tr) and CMVPV, \MB{for observed locations ($\mathscr{L}_O$) only} denoted collectively here by \MB{$v_o$}. We examine the following cutoff options: 
\begin{enumerate}
    \item Maximum predictive variance (Cook, Conn) 
    \begin{equation}
        k_{\text{max}} = \max_{\MB{o} \in \mathscr{L}_O}(v_\MB{o})
    \end{equation}
    \item Leverage-informed maximum predictive variance
    $$k_{\text{lev}} = \max_{\MB{o} \in \mathscr{L}_{O,-\text{lev}}}(v_\MB{o})$$
    \item Quantile value
    $$k_r = q_r (v_\MB{o}, \MB{o} \in  \mathscr{L}_O)$$
  
\end{enumerate}
The leverage-informed cutoff value is calculated from a set of observations in $\mathscr{L}_{O, -\text{lev}}$, where potential influential points (as determined based only on the covariate values, $\mathbf{X}$) have been removed. We suggest considering quantile-based approaches as cutoff values at the 0.99 and 0.95 quantiles of the prediction variances from observed locations. These cutoffs are less conservative than the maximum predictive variance which may also be considered the 100\% quantile value (i.e. a smaller cutoff value results in more unobserved locations identified as places where the empirical model may not be trusted).

\subsection*{Identifying \MB{l}ocations as \MB{e}xtrapolations}

With the (C)MVPV values and cutoff choice in hand, determining which locations (observed/unobserved) are extrapolations is straightforward and results in a binary (yes/no) value. We refer to this delineation as our extrapolation index (e)  
\begin{equation}
    \text{e\textsuperscript{k}}_\MB{p} = \left\{ 
    \begin{array}{cc}
        1 &  \text{ if } \MB{v_p} > k\\
        0 & \text{otherwise}
    \end{array}
     \right.
\end{equation} 
where $k$ represents the cutoff choice \MB{obtained using $v_o$}. \MB{Each extrapolation index value is a function of the scalar value prediction variance representations of MVPV(D or tr) and CMVPV, for predicted locations ($\mathscr{L}_P$) only denoted collectively here by $v_p$.} While this binary formulation allows for a simple way to determine whether or not we may diagnose a point as being an extrapolation, it does not allow for much nuance. Should a prediction with its predictive variance just beyond the boundary of the IVH be considered as untrustworthy as one with a predictive variance well beyond the boundary? We thus propose a numeric measure of extrapolation calculated by dividing predictive variance values \MB{for predicted locations} by the cutoff value to generate a Relative MVPV (RMVPV) measurement:
\begin{equation}
    \text{R\textsubscript{k}MVPV\textsubscript{\MB{p}}} = \frac{ \MB{v_p}}{k}.
\end{equation}
R\textsubscript{k}MVPV values greater than 1 would be considered to be extrapolations, but in addition the larger the value the less trustworthy we would consider its prediction to be. \MB{This approach does not change which locations are identified as extrapolations since the binary} extrapolation index \MB{as described above} can be calculated from the RMVPV as
\begin{equation}
    \text{e\textsuperscript{k}}_\MB{p} = \left\{ 
    \begin{array}{cc}
        1 &  \text{ if R\textsubscript{k}MVPV\textsubscript{\MB{p}}} > 1\\
        0 & \text{otherwise}
    \end{array}
     \right..
\end{equation}

\subsection*{Choosing IVH vs PV}

With several methods of identifying extrapolations available we now provide additional guidance on choosing between various options. Cook's approach of using the maximum leverage value to define the IVH boundary may be useful for either an univariate or a joint model in a linear regression framework. However, as it depends on covariate values alone, it lacks any influence of response data. Conn et al.'s  gIVH introduces the use of posterior predictive variance instead of the hat matrix to define the hull boundary in the case of a generalized model. 

One possible limitation of predictive variance approaches to obtain an extrapolation index arises under certain generalized models. Models with constrained supports (i.e. binary, Poisson, etc) may exhibit decreased posterior variation when predictions are near the edges of the support. For example, in the binary case with a single covariate, if 
\begin{linenomath}
\begin{equation}
    \begin{aligned}
    y_i &\sim \text{Bern}(p_i)\\
    \text{logit}(p_i) &= \beta_0 + \beta_1 x_i
\end{aligned}
\end{equation}
 \end{linenomath}
then as $x_i  \rightarrow \infty \: (x_i \leftarrow -\infty), p_i \rightarrow 1 \: (p_i \leftarrow 0),$ and $\text{var}(y_i|p_i) = p_i(1-p_i) \rightarrow 0$. Thus, extreme points on the outside range of the observed values may have tiny predicted variance. This artificial decrease in variance may mask the identification of potentially extrapolated data points when using PV methods. Missing these extrapolations may also hinder our ability to characterize the covariate space, limiting the ability to provide reliable predictions. Thus, in models where prediction variance decreases as means go to extreme value\MB{s such as Binomial, Beta, or Uniform distributions}, we recommend IVH over PV approaches \MB{where this masking of extrapolation locations does not occur}.
We use the inland lake data set (see~\nameref{s:LAGOS}) to illustrate predicting joint response variables at unobserved lake locations. 

\subsection*{Visualization and \MB{i}nterpretation}

Exploring data and taking a principled approach to identifying potential extrapolation points is often aided by visualization (and interpretation) of data and predictions. With the LAGOS data we examine spatial plots of the lakes and their locations coded by extrapolation vs prediction. Plotting this for multiple cutoff choices (as in Fig \ref{f:tracePV}) is useful to explore how this choice can influence which locations are considered extrapolations. This is important from both an ecological and management perspective. For instance, if potential areas are identified as having many extrapolations this might suggest that specific lake ecosystems or landscapes have characteristics influencing processes governing nutrient dynamics in lakes not well captured by previously collected data -- and thus may require further investigation.

    In addition to an exploration of possible extrapolation in physical space (through the plot in Fig \ref{f:tracePV}), we also examine possible extrapolation in covariate space. Using either of the binary/numeric Extrapolation Index values, we propose a Classification and Regression Tree (CART) analysis with the extrapolation values as the response. Our classification approach allows for further insight into what covariates may be influential in determining whether a newly observed location is too dissimilar to existing ones. A CART model allows for the identification of regions in covariate \EH{space} where predictions are suspect and may inform future sampling efforts as the available data has not fully characterized all lakes.

\subsection*{Model \MB{f}itting}
The joint nutrient-productivity model (see~\nameref{s:MVcase}) was fit using MCMC in R \cite{RCoreTeam2016}. We ran the MCMC algorithm for 20,000 iterations and used the \textbf{coda} package to analyze MCMC output and check for convergence \cite{Plummer2006}. Full conditional updates were available for all parameters ($\mathbf{B}, \bm{\Sigma}$, and $\mathbf{Z}$) thus Gibbs updates were specified. We generated posterior predictions of lake nutrient levels across the entirety of observed and unobserved lake locations as
\begin{equation}
    \mathbf{y}_{i} \sim \text{N}(\mathbf{Bx}_{\MB{i}}, \bm{\Sigma}) \quad i = 1, \dots, n
\end{equation}
and calculated multivariate prediction variance values as described in \nameref{ss:MVPV}.

\section*{Results}
\begin{figure}[h]
  \centerline{\includegraphics[width=\textwidth]{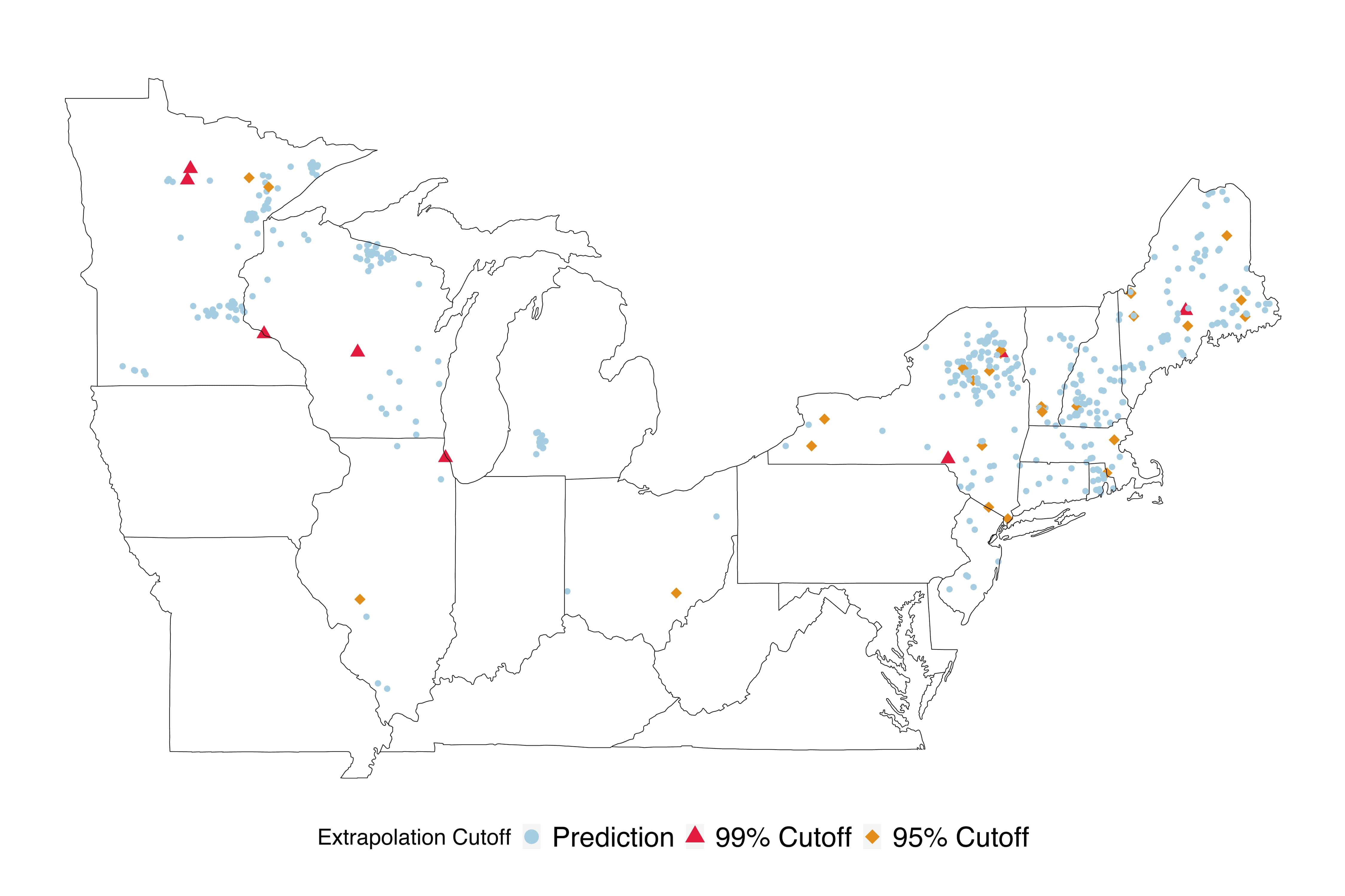}}
\caption{Identification of prediction vs extrapolation locations of LAGOS-NE lakes. Four cutoff approaches are compared and presented. Lakes in \MB{orange diamonds and} red \MB{triangles} indicate those where predictions were beyond the \MB{99\% and 95\%} cutoff value\MB{s, respectively,} and thus considered extrapolations. \MB{The color and shape of extrapolated lake locations are determined by which cutoff value first identifies the prediction at that location as an extrapolation.}}
\label{f:tracePV}
\end{figure}

\MB{Fitting our multivariate linear model to the 8,910 lakes resulted in most lakes' predictions} remaining within the extrapolation index cutoff and thus not being identified as extrapolations. We explored the use of both trace and determinant for obtaining a scalar value representation of the multivariate posterior predictive variance in addition to four cutoff criteria. \MB{Using MVPV(tr) with t}hese cutoffs (max value, leverage max, 0.99 quantile, and 0.95 quantile) resulted in 0, 1, 9, and 33  multivariate response predictions being identified as extrapolations, respectively. \MB{\TW{In contrast,} using MVPV(D) values combined with the four cutoffs resulted in 0, 0, 8, 37 predictions identified as extrapolations. Unless all response variables are on the same scale we recommend the use of MVPV(D) over MVPV(tr). However, if a scale-invariant measure if not necessary, exploring the use of MVPV(tr) (in addition to MVPV(D) may reveal single-response variables that are of interest to researchers for further exploration using our Conditional MVPV approach.}
Fig \ref{f:tracePV} shows the spatial locations of lakes where the collective model predictions for TP, TN, Chl a, and Secchi depth have been identified as extrapolations  \MB{using MVPV(D) combined with the} the cutoff measures. As the cutoff values become more conservative in nature the number of extrapolations identified increases. \MB{This figure shows the level of cutoff that first identifies a location as an extrapolation, (e.g. red squares are locations first flagged using the 99\% cutoff, but they would also be included in the extrapolations found with the 95\% cutoff).} This increasing number of extrapolations identified highlights the importance of exploring different choices for a cutoff value. When the maximum value \MB{or the leverage-informed maximum} of the predictive variance measure ($k_{\text{max}}$ \MB{and $k_{\text{lev}}$}) \MB{are} used as cutoff\MB{s} for determining when a prediction for an unsampled lake location should not be fully trusted, zero lakes are identified as extrapolations. 

\begin{figure}[h]
  \centerline{\includegraphics[width=.7\textwidth]{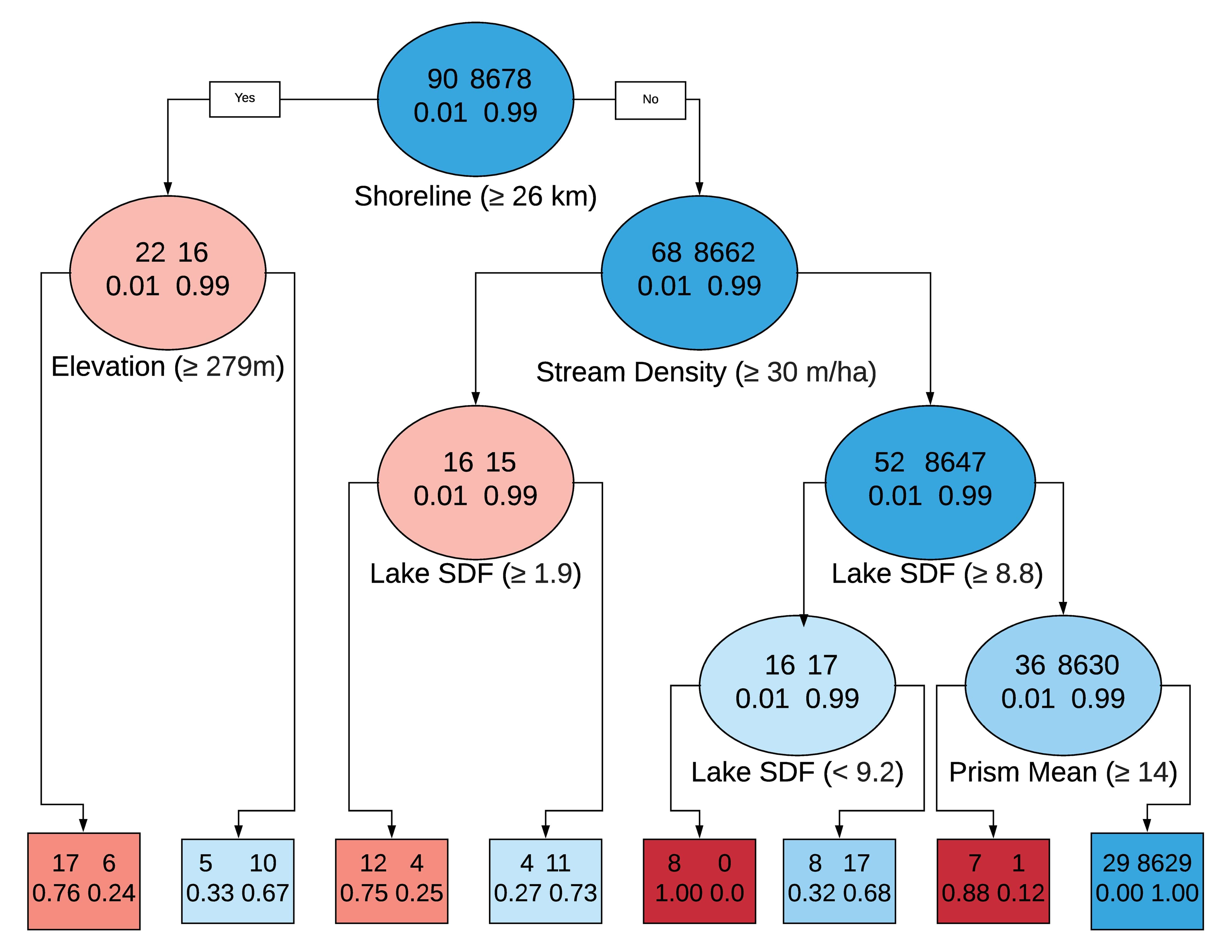}}
\caption{CART model results showing which variables may be useful in identifying extrapolations for inland lake. \MB{Each level of nodes include the thresholds and variables used to sort the data. Node color indicate whether the majority of sorted inland lake locations were identified as predictions (blue) or extrapolations (red). The first row of numbers in a node indicate the number of lakes identified as predictions (right) or extrapolations (left) that have been sorted into this node. The second row of numbers indicate the percentage of lakes that are identified as predictions (left) or extrapolation (right) with the terminal nodes (square nodes) including the percentage of records sorted by the decision tree.}}
\label{f:CART95dMV}
\end{figure}

Exploratory data analysis (see \nameref{S1_Fig}) indicates that for each of the lakes identified as extrapolations, the values are within the distribution of the data, with only a few exceptions. Rather than a few key variables standing out, it appears to be some combination of variables that makes a lake an extrapolation. To further characterize the type of lake  more likely to be identified as an extrapolation we used a CART Model with our binary extrapolation index results using the MVPV(D) and the 0.95 quantile cutoff. This approach can help identify regions in the covariate space where extrapolations are more likely to occur (Fig \ref{f:CART95dMV}). This CART analysis suggests the most important factors associated with extrapolation include shoreline length, elevation, stream density, and lake SDF. For example, a lake with a shoreline greater than 26 kilometers and above a certain elevation ($\geq$ 279 m), is likely to be identified as an extrapolation when using this model to obtain predictions. 
This type of information is useful for ecologists trying to model lake nutrients because it suggests lakes with these types of characteristics may behave differently than other lakes. In fact, lake perimeter, SDF, and elevation have been shown to be associated with reservoirs relative to natural lakes \cite{Doubek2017CatchmentStates}. Although it is beyond the scope of our paper to fully explore this \TW{notion} because our existing database does not differentiate between natural lakes and reservoirs, these results lend support to our approach and conclusions.

\begin{figure}[h]
   \centerline{\includegraphics[width=\textwidth]{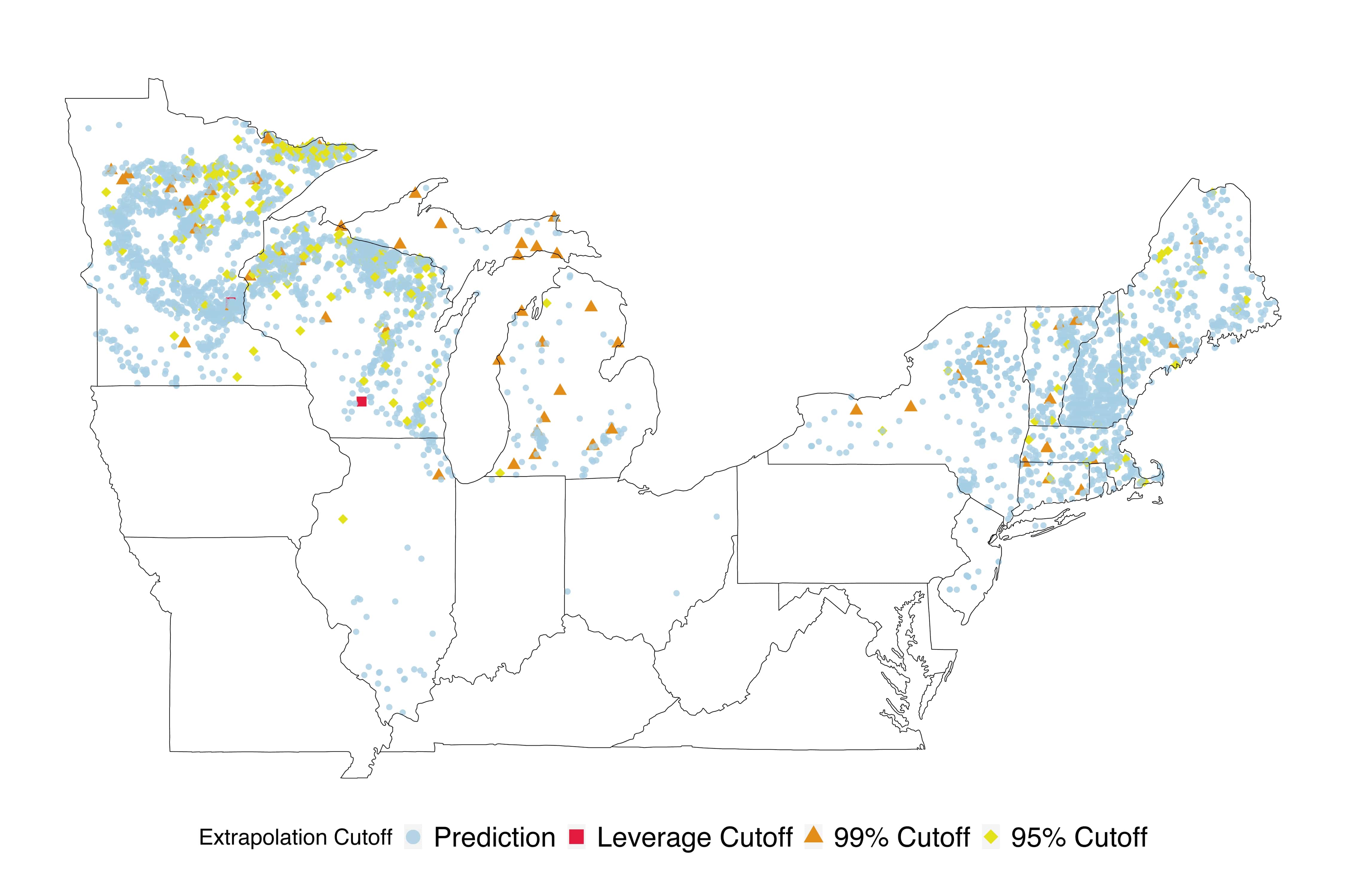}}
\caption{Identification and locations of prediction vs extrapolation of the single response variable, TN. Four cutoff approaches are compared and presented. Lakes in \MB{blue circles represent locations where TN predictions have been not been identified as extrapolations for any cutoff choice. Lakes  in}red\MB{squares, orange triangles, and yellow diamonds} indicate those where predictions were beyond the cutoff value\MB{s} and thus considered extrapolations. \MB{The color and shape of extrapolated lake locations are determined by which cutoff value first identifies the prediction at that location as an extrapolation.}}
\label{f:TNonlyCMVPV}
\end{figure}

We also employed the conditional single variable extrapolation through predictive variance approach to leverage all information known about a lake when considering whether a prediction of a single response variable (e.g. TN, as explored here) is an extrapolation (Fig \ref{f:TNonlyCMVPV}). These cutoffs resulted in  0, 2, 73, and 386 lake multivariate response predictions out of 5031 being identified as extrapolations. To characterize the type of lake more likely to be identified as an extrapolation we used a CART model using the 95\% cutoff criterion. CART revealed the most important factors associated with extrapolation were latitude, maximum depth, and watershed to lake size ratio. Latitude may be expected as many of the lakes without measures for TN are located in the northern region. An additional visualization and table exploring extrapolation lakes and their covariate values may be found in \nameref{S2_Table}.  

\section*{Discussion}
We have presented different approaches for identifying and characterizing potential extrapolation points within multivariate response data. Ecological research is often faced with the challenge of explaining processes at broad scales with limited data. Financial, temporal, and logistical restrictions often prevent research efforts from fully exploring an ecosystem or ecological setting. Rather, ecologists rely on predictions made on a select amount of available data that may not fully represent the  breadth of a system of study. By better understanding when extrapolation is occurring scientists may avoid making unsound inferences.

In our inland lakes example we addressed the issue of large-scale predictions to fill in missing data using a joint linear model presented by Wagner and Schliep \cite{Wagner2018}. With our novel approach for identifying and characterizing extrapolation in a multivariate setting we were able to provide numeric measures associated with extrapolation (MVPV, CMVPV, R(C)MVPV) allowing for focus on predictions for all response variables or a single response variable while conditioning on others. Each of these measurements, when paired with a cutoff criterion, identify novel locations that are extrapolations. Our recommendations for visualization and interpretation of these extrapolated lakes is useful for future analyses and predictions which inform policy and management decisions. Insight into identified extrapolations and their characteristics provides additional sampling location\MB{s to consider} for future work. In this analysis we found  certain lakes, such as lakes located at relatively higher elevations in our study area, are more likely to be identified as an extrapolation. The available data may thus not fully represent these types of lakes, resulting in them being poorly predicted, or identified as extrapolations. 

The tools outlined in this work provide novel insights into identifying and characterizing extrapolations in multivariate response settings. \MB{Further extensions of this work are available but not explored in this paper. In addition to the A- and D-optimality approaches (trace and determinant, respectively) used to obtain scalar value representations of the covariance matrices one may also explore the utility of E-optimality (maximum eigenvalue) as an additional criterion. This approach would focus on examining variance in the first principle component of the predictive variance matrix and, like the trace,  this variance is not a scale-invariant measure.} Our work takes advantage of posterior predictive inference \MB{under a Bayesian setting} to obtain an estimate of the variance of the predictive mean response \MB{vector for each lake}. However, \MB{a frequentist approach using }simulation-based methods \MB{may also} provide an estimate of this variance through \MB{non-parametric or parametric} bootstrapping \MB{(a comparison of the two for spatial abundance estimates may be found in Hedley and Buckland \cite{Hedley2004SpatialSampling})} and the extrapolation coefficients may be obtained through the trace and/or determinant of this variance. 

This work results in identification of extrapolated lake locations as well as further understanding of the unique covariate space they occupy. The resulting caution shown when using joint nutrient models to estimate water quality variables at lakes with partially or completely unsampled measures is necessary for larger goals such as estimating the overall combined levels of varying water qualities in all US inland lakes. In addition, under- or overestimating concentrations of key nutrients such as TN and TP can potentially
lead to misinformed management strategies which may have deleterious effects on water quality and the lake ecosystem. \TW{T}he identification of lake and landscape characteristics associated with extrapolation locations can further understanding between natural/anthropogenic sources of nutrients in lakes not well represented in the sampled population. In our database, TP is sampled more than TN, which is likely due to the conventional wisdom that inland waters are P limited  \MB{,where P} contribute\MB{s} the most to eutrophication \cite{Conley2009EcologyPhosphorus}. However, nitrogen has been shown to be an important nutrient in eutrophication in some lakes and some regions \cite{Paerl2011ControllingStrategy}, and may be as important to sample to fully understand lake eutrophication. Our results show it is possible to predict TN if other water quality variables are available, but it would be better if it was sampled more often. 

The joint model used in this work can be improved upon in several regards; no spatial component is included, response variables are averages over several years worth of data and thus temporal variation is not considered, and data from different years are given  equal weight. The model we use to fit these data may be considered to be a simple one, but the novel approach presented here may be applied to more complicated models. In a sample based approach using a Bayesian framework the MVPV and CMVPV values obtained come from the MCMC samples and are thus independent from model design choices.

Deeper understanding of where extrapolation is occurring will allow researchers to propagate this uncertainty forward. Follow up analyses using model-based predictions need to acknowledge that some predictions are less trustworthy than others. This approach and our analysis here shows that while a model may be able to produce an estimate and a confidence or prediction interval, that does not mean the truth is captured nor does the assumed relationship persist, especially outside the range of observed data. The methods outlined here will serve to guide future scientific inquiries involving joint distribution models.

\section*{Supporting information}

\paragraph*{S1 Fig.}
\label{S1_Fig}
{\bf Violin plots of covariate densities and extrapolation points plotted.}

\paragraph*{S2 Tables}
\label{S2_Table}
{\bf Tables of covariate values for lakes identified as extrapolations using MVPV(\MB{D}) and CMVPV for TN.}

\section*{Acknowledgments}

We thank the LAGOS Continental Limnology Research Team for helpful discussions throughout the process of this manuscript. Any use of trade, firm, or product names is for descriptive purposes only and does not imply endorsement by the U.S. Government.

\nolinenumbers

 \end{document}


\begin{flushleft}
{\Large
\textbf\newline{\MB{S2 Tables: Tables of covariate values for lakes identified as extrapolations using MVPV(D) and CMVPV for TN.  }}
}
\newline
\end{flushleft}

\MB{In this paper, we present new methods for identifying extrapolation within a multivariate setting. This novel work provides numeric measures associated with extrapolation (MVPV, CMVPV, R(C)MVPV) allowing for focus on predictions for all response variables or a single response variable while conditioning on others. In S1 Figures we visualized a selection of the lakes identified as extrapolations to examine whether the covariate values for each location are within the distribution of the data. Tables \ref{t:mvpvD} and \ref{t:Cmvpv} included below contain the values of each covariate included in model described in Extension to to Multivariate Case.}



 \begin{table}[ht]
\centering
\caption{\MB{Covariate values for out-of-sample lakes where the determinant of multivariate predictive variance matrices for each inland lake (MVPV(D)) exceeds a 95\% quantile cutoff value obtained using MVPV(D) values from in-sample lakes.}}
\label{t:mvpvD}
\resizebox{\textwidth}{!}{%
\begin{tabular}{r|rrrrrrrrrrrrrrrrr}
  \hline
  \rot{Lake ID} & \rot{maxdepth\_m} & \rot{IWS\_lk\_ratio} & \rot{iws\_nlcd2006\_for} & \rot{lake\_sdf} & \rot{streamdensity\_mperha} & \rot{elevation\_m} & \rot{wlconnections\_shoreline\_km} & \rot{buff\_roaddensity\_mperha} & \rot{baseflow\_mean} & \rot{runoff\_mean} & \rot{hu4\_nlcd2006\_agr} & \rot{totalNdep\_1990\_mean} & \rot{totalNdep\_2010\_mean} & \rot{prism\_ppt\_mean} & \rot{prism\_tmean\_mean} & \rot{DR\_Stream} & \rot{HW\_ISO} \\ 
   \hline
41210 & 21.30 & 1.16 & 50.85 & 1.63 & 3.43 & 396.16 & 42.31 & 13.16 & 58.44 & 5.83 & 34.03 & 4.37 & 4.48 & 713.97 & 5.44 & 0.00 & 0.00 \\ 
  25421 & 13.41 & 7.56 & 34.43 & 4.76 & 4.48 & 281.32 & 42.70 & 30.63 & 60.20 & 10.41 & 34.07 & 5.75 & 4.73 & 839.27 & 6.47 & 0.00 & 0.00 \\ 
  25741 & 35.10 & 0.66 & 55.48 & 2.63 & 1.74 & 396.77 & 30.17 & 41.30 & 58.44 & 5.83 & 34.03 & 4.37 & 4.48 & 713.97 & 5.44 & 0.00 & 0.00 \\ 
  29301 & 17.10 & 30.47 & 19.92 & 3.67 & 12.00 & 203.00 & 38.45 & 81.27 & 58.38 & 7.26 & 52.17 & 6.25 & 5.41 & 850.50 & 7.09 & 0.00 & 0.00 \\ 
  5489 & 198.40 & 6.70 & 30.37 & 2.72 & 12.52 & 135.43 & 0.06 & 102.06 & 54.55 & 19.15 & 36.29 & 8.67 & 4.01 & 1037.07 & 8.17 & 0.00 & 0.00 \\ 
  5804 & 24.38 & 15.54 & 63.99 & 11.07 & 14.74 & 128.03 & 2.04 & 8.46 & 19.89 & 10.70 & 47.29 & 5.88 & 4.02 & 1106.82 & 12.98 & 1.00 & 0.00 \\ 
  66251 & 7.01 & 19.04 & 57.79 & 1.85 & 39.51 & 83.32 & 1.01 & 31.44 & 47.76 & 25.27 & 3.96 & 4.53 & 2.09 & 1146.49 & 5.02 & 0.00 & 0.00 \\ 
  75271 & 57.00 & 37.45 & 95.92 & 3.48 & 19.53 & 438.74 & 0.06 & 15.89 & 52.11 & 22.77 & 16.41 & 6.56 & 4.09 & 1185.83 & 10.48 & 0.00 & 1.00 \\ 
  10315 & 68.20 & 35.51 & 86.74 & 6.02 & 14.09 & 389.95 & 1.45 & 78.66 & 52.11 & 22.77 & 16.41 & 6.56 & 4.09 & 1185.83 & 10.48 & 0.00 & 0.00 \\ 
   \hline
\end{tabular}
}
\end{table}

\begin{table}[ht]
\centering
\caption{\MB{Covariate values for out-of-sample lakes where the multivariate predictive variance scalar values for each inland lake's TN response variable (CMVPV) exceeds a 95\% quantile cutoff value obtained using CMVPV values from in-sample lakes.}}
\label{t:Cmvpv}
\resizebox{.8\textwidth}{!}{%
\begin{tabular}{r|rrrrrrrrrrrrrrrrr}
  \hline
 \rot{Lake ID} & \rot{maxdepth\_m} & \rot{IWS\_lk\_ratio} & \rot{iws\_nlcd2006\_for} & \rot{lake\_sdf} & \rot{streamdensity\_mperha} & \rot{elevation\_m} & \rot{wlconnections\_shoreline\_km} & \rot{buff\_roaddensity\_mperha} & \rot{baseflow\_mean} & \rot{runoff\_mean} & \rot{hu4\_nlcd2006\_agr} & \rot{totalNdep\_1990\_mean} & \rot{totalNdep\_2010\_mean} & \rot{prism\_ppt\_mean} & \rot{prism\_tmean\_mean} & \rot{DR\_Stream} & \rot{HW\_ISO} \\ 
    \hline
102 & 14.94 & 5.99 & 68.25 & 1.84 & 0.00 & 226.84 & 1.50 & 11.27 & 65.90 & 11.40 & 53.80 & 7.38 & 4.66 & 928.54 & 9.05 & 0.00 & 1.00 \\ 
  242 & 2.80 & 21.01 & 15.72 & 1.82 & 8.96 & 278.95 & 0.00 & 94.20 & 58.44 & 5.83 & 34.03 & 4.37 & 4.48 & 713.97 & 5.44 & 1.00 & 0.00 \\ 
  462 & 18.00 & 2.39 & 72.96 & 2.04 & 0.00 & 427.22 & 0.07 & 0.00 & 50.38 & 8.03 & 2.60 & 2.92 & 2.90 & 689.37 & 3.46 & 0.00 & 1.00 \\ 
  479 & 1.98 & 92.16 & 47.27 & 2.21 & 4.57 & 197.24 & 2.51 & 3.89 & 62.10 & 17.22 & 1.94 & 4.18 & 3.14 & 818.74 & 4.94 & 0.00 & 0.00 \\ 
  530 & 3.35 & 1.24 & 17.80 & 1.40 & 1.28 & 259.55 & 0.00 & 92.77 & 65.90 & 11.40 & 53.80 & 7.38 & 4.66 & 928.54 & 9.05 & 0.00 & 0.00 \\ 
  558 & 1.70 & 1.59 & 3.53 & 4.23 & 7.14 & 298.48 & 16.01 & 15.74 & 43.83 & 3.34 & 77.69 & 4.92 & 6.15 & 712.47 & 6.89 & 0.00 & 0.00 \\ 
  889 & 3.66 & 1937.44 & 60.78 & 3.56 & 9.63 & 299.28 & 0.66 & 3.13 & 52.26 & 12.14 & 3.99 & 4.23 & 3.35 & 781.28 & 4.04 & 0.00 & 0.00 \\ 
  906 & 3.35 & 4.10 & 61.56 & 1.18 & 0.00 & 538.01 & 0.41 & 2.95 & 60.21 & 11.29 & 27.98 & 5.36 & 3.89 & 790.51 & 6.11 & 0.00 & 1.00 \\ 
  932 & 3.35 & 44.06 & 42.98 & 1.59 & 10.21 & 191.08 & 1.93 & 9.78 & 78.83 & 14.64 & 13.13 & 5.76 & 3.62 & 848.36 & 6.84 & 0.00 & 1.00 \\ 
  962 & 3.96 & 2.87 & 71.77 & 1.26 & 0.03 & 524.02 & 0.95 & 33.23 & 60.21 & 11.29 & 27.98 & 5.36 & 3.89 & 790.51 & 6.11 & 0.00 & 1.00 \\ 
  982 & 1.80 & 10.22 & 42.89 & 3.01 & 3.40 & 402.30 & 26.01 & 3.26 & 50.38 & 8.03 & 2.60 & 2.92 & 2.90 & 689.37 & 3.46 & 0.00 & 0.00 \\ 
  1053 & 3.96 & 343.96 & 32.13 & 3.85 & 10.15 & 176.99 & 5.14 & 53.97 & 78.83 & 14.64 & 13.13 & 5.76 & 3.62 & 848.36 & 6.84 & 0.00 & 0.00 \\ 
  1136 & 3.66 & 48.39 & 38.07 & 1.87 & 7.43 & 186.56 & 5.04 & 75.22 & 78.83 & 14.64 & 13.13 & 5.76 & 3.62 & 848.36 & 6.84 & 1.00 & 0.00 \\ 
  1158 & 7.00 & 5.04 & 52.34 & 5.13 & 2.08 & 561.85 & 0.52 & 0.00 & 52.26 & 12.14 & 3.99 & 4.23 & 3.35 & 781.28 & 4.04 & 0.00 & 0.00 \\ 
  1311 & 3.05 & 7.84 & 32.70 & 1.53 & 2.03 & 184.71 & 4.37 & 48.89 & 78.83 & 14.64 & 13.13 & 5.76 & 3.62 & 848.36 & 6.84 & 1.00 & 0.00 \\ 
  1336 & 2.40 & 10.25 & 63.26 & 2.05 & 0.01 & 450.14 & 1.12 & 50.70 & 50.38 & 8.03 & 2.60 & 2.92 & 2.90 & 689.37 & 3.46 & 0.00 & 1.00 \\ 
  1353 & 3.66 & 7444.23 & 34.44 & 4.17 & 10.98 & 225.08 & 3.92 & 1.21 & 60.20 & 10.41 & 34.07 & 5.75 & 4.73 & 839.27 & 6.47 & 0.00 & 0.00 \\ 
  1361 & 17.68 & 9.40 & 57.13 & 1.37 & 0.00 & 332.83 & 0.72 & 0.13 & 60.21 & 11.29 & 27.98 & 5.36 & 3.89 & 790.51 & 6.11 & 0.00 & 1.00 \\ 
  1386 & 2.13 & 4.63 & 34.95 & 1.49 & 3.01 & 214.89 & 3.65 & 16.16 & 78.83 & 14.64 & 13.13 & 5.76 & 3.62 & 848.36 & 6.84 & 1.00 & 0.00 \\ 
  1532 & 1.83 & 25.75 & 39.01 & 1.52 & 5.30 & 323.99 & 1.43 & 5.84 & 62.10 & 17.22 & 1.94 & 4.18 & 3.14 & 818.74 & 4.94 & 1.00 & 0.00 \\ 
  1656 & 1.83 & 15.89 & 30.41 & 2.80 & 0.02 & 235.79 & 0.00 & 79.25 & 65.90 & 11.40 & 53.80 & 7.38 & 4.66 & 928.54 & 9.05 & 0.00 & 0.00 \\ 
  1661 & 28.35 & 225.83 & 88.38 & 2.12 & 13.46 & 372.91 & 0.53 & 5.34 & 51.67 & 23.34 & 13.99 & 6.28 & 3.12 & 1101.03 & 6.15 & 1.00 & 0.00 \\ 
  1714 & 2.13 & 4.10 & 58.81 & 1.42 & 5.29 & 289.39 & 0.97 & 63.89 & 65.90 & 11.40 & 53.80 & 7.38 & 4.66 & 928.54 & 9.05 & 0.00 & 0.00 \\ 
  2059 & 2.44 & 16.67 & 68.67 & 2.99 & 0.00 & 379.40 & 1.44 & 48.61 & 78.83 & 14.64 & 13.13 & 5.76 & 3.62 & 848.36 & 6.84 & 0.00 & 1.00 \\ 
  2203 & 17.40 & 4.97 & 51.19 & 2.04 & 4.69 & 397.17 & 6.90 & 21.69 & 58.44 & 5.83 & 34.03 & 4.37 & 4.48 & 713.97 & 5.44 & 0.00 & 0.00 \\ 
  2243 & 3.05 & 1.89 & 81.96 & 1.44 & 0.00 & 529.98 & 0.64 & 8.05 & 60.21 & 11.29 & 27.98 & 5.36 & 3.89 & 790.51 & 6.11 & 0.00 & 1.00 \\ 
  2323 & 2.44 & 10.09 & 97.44 & 1.79 & 0.85 & 453.89 & 0.00 & 9.47 & 47.92 & 24.18 & 6.59 & 5.88 & 2.87 & 1205.86 & 6.80 & 1.00 & 0.00 \\ 
  2424 & 13.40 & 3.03 & 46.01 & 1.98 & 0.00 & 419.64 & 0.15 & 113.22 & 52.26 & 12.14 & 3.99 & 4.23 & 3.35 & 781.28 & 4.04 & 0.00 & 1.00 \\ 
  2435 & 32.00 & 4.30 & 32.88 & 4.55 & 6.57 & 365.95 & 11.56 & 48.41 & 58.44 & 5.83 & 34.03 & 4.37 & 4.48 & 713.97 & 5.44 & 0.00 & 0.00 \\ 
  2507 & 1.83 & 44.53 & 56.35 & 2.02 & 7.77 & 175.50 & 8.91 & 31.89 & 50.12 & 9.05 & 47.43 & 5.75 & 3.89 & 815.29 & 8.04 & 0.00 & 0.00 \\ 
  2784 & 2.74 & 15.59 & 30.70 & 3.60 & 2.41 & 266.89 & 1.58 & 67.64 & 73.63 & 13.12 & 7.12 & 5.13 & 3.44 & 786.90 & 6.12 & 0.00 & 0.00 \\ 
  2785 & 4.27 & 5.59 & 5.90 & 1.21 & 1.96 & 249.50 & 3.47 & 11.66 & 65.90 & 11.40 & 53.80 & 7.38 & 4.66 & 928.54 & 9.05 & 1.00 & 0.00 \\ 
  2946 & 1.83 & 19.42 & 63.64 & 1.24 & 3.09 & 187.96 & 0.32 & 43.78 & 78.83 & 14.64 & 13.13 & 5.76 & 3.62 & 848.36 & 6.84 & 1.00 & 0.00 \\ 
  3115 & 2.44 & 3.09 & 54.52 & 2.80 & 0.00 & 292.55 & 0.56 & 6.11 & 65.90 & 11.40 & 53.80 & 7.38 & 4.66 & 928.54 & 9.05 & 0.00 & 1.00 \\ 
  3360 & 2.13 & 6.46 & 19.28 & 1.40 & 0.00 & 269.30 & 0.48 & 100.21 & 50.12 & 9.05 & 47.43 & 5.75 & 3.89 & 815.29 & 8.04 & 0.00 & 1.00 \\ 
  3694 & 10.97 & 19.40 & 14.27 & 1.92 & 1.01 & 286.83 & 0.00 & 22.67 & 57.50 & 9.75 & 24.44 & 5.44 & 4.78 & 792.99 & 5.69 & 0.00 & 0.00 \\ 
  3848 & 14.02 & 6.84 & 80.97 & 1.67 & 9.67 & 392.17 & 1.07 & 84.66 & 55.01 & 10.60 & 26.33 & 5.82 & 4.65 & 821.25 & 5.70 & 0.00 & 0.00 \\ 
  4100 & 4.88 & 189.50 & 31.99 & 4.39 & 11.70 & 316.08 & 2.70 & 63.17 & 55.01 & 10.60 & 26.33 & 5.82 & 4.65 & 821.25 & 5.70 & 1.00 & 0.00 \\ 
  4152 & 8.23 & 4.50 & 74.99 & 1.23 & 0.00 & 308.50 & 0.72 & 17.71 & 57.50 & 9.75 & 24.44 & 5.44 & 4.78 & 792.99 & 5.69 & 0.00 & 1.00 \\ 
  4400 & 9.45 & 10.52 & 85.89 & 1.89 & 23.00 & 502.90 & 1.24 & 2.39 & 51.67 & 23.34 & 13.99 & 6.28 & 3.12 & 1101.03 & 6.15 & 1.00 & 0.00 \\ 
  5341 & 5.50 & 15.50 & 42.75 & 1.34 & 8.87 & 93.30 & 0.27 & 57.72 & 47.92 & 24.18 & 6.59 & 5.88 & 2.87 & 1205.86 & 6.80 & 1.00 & 0.00 \\ 
  5373 & 24.40 & 3.63 & 56.65 & 1.59 & 2.76 & 222.60 & 1.14 & 92.06 & 49.72 & 25.12 & 8.35 & 6.58 & 3.16 & 1247.05 & 9.39 & 1.00 & 0.00 \\ 
  5693 & 1.83 & 7.70 & 0.00 & 1.64 & 0.39 & 215.11 & 0.00 & 166.24 & 41.86 & 10.64 & 60.78 & 7.52 & 5.21 & 954.63 & 9.75 & 0.00 & 1.00 \\ 
  6077 & 15.24 & 2.34 & 87.03 & 1.41 & 0.00 & 318.02 & 0.19 & 20.82 & 50.29 & 26.02 & 1.79 & 4.27 & 2.01 & 1130.34 & 4.96 & 0.00 & 1.00 \\ 
  6517 & 4.57 & 23.18 & 67.04 & 1.22 & 17.76 & 52.65 & 1.74 & 0.00 & 47.76 & 25.27 & 3.96 & 4.53 & 2.09 & 1146.49 & 5.02 & 1.00 & 0.00 \\ 
  6725 & 1.60 & 585.63 & 68.79 & 2.53 & 12.06 & 99.28 & 0.65 & 35.03 & 57.05 & 24.67 & 4.80 & 4.91 & 3.11 & 1241.95 & 10.02 & 0.00 & 0.00 \\ 
  6939 & 1.79 & 15.56 & 18.00 & 1.32 & 0.00 & 266.75 & 1.79 & 0.00 & 58.44 & 5.83 & 34.03 & 4.37 & 4.48 & 713.97 & 5.44 & 0.00 & 1.00 \\ 
  7090 & 23.80 & 3.38 & 48.51 & 1.38 & 0.12 & 396.18 & 1.97 & 0.00 & 58.44 & 5.83 & 34.03 & 4.37 & 4.48 & 713.97 & 5.44 & 0.00 & 1.00 \\ 
  7225 & 18.90 & 9.89 & 78.65 & 2.86 & 8.10 & 478.95 & 0.57 & 103.54 & 51.67 & 23.34 & 13.99 & 6.28 & 3.12 & 1101.03 & 6.15 & 0.00 & 0.00 \\ 
  7305 & 135.00 & 1.83 & 42.14 & 2.36 & 0.00 & 440.82 & 0.00 & 27.20 & 52.26 & 12.14 & 3.99 & 4.23 & 3.35 & 781.28 & 4.04 & 0.00 & 1.00 \\ 
  7349 & 1.94 & 171.66 & 5.86 & 1.41 & 14.07 & 266.45 & 0.74 & 12.72 & 58.44 & 5.83 & 34.03 & 4.37 & 4.48 & 713.97 & 5.44 & 1.00 & 0.00 \\ 
  7485 & 6.00 & 53517.40 & 15.43 & 1.20 & 7.97 & 272.85 & 0.78 & 49.54 & 58.44 & 5.83 & 34.03 & 4.37 & 4.48 & 713.97 & 5.44 & 0.00 & 0.00 \\ 
  7541 & 15.50 & 3.88 & 87.95 & 1.68 & 10.86 & 522.94 & 0.00 & 58.53 & 51.67 & 23.34 & 13.99 & 6.28 & 3.12 & 1101.03 & 6.15 & 1.00 & 0.00 \\ 
  7732 & 2.00 & 19.76 & 6.80 & 1.17 & 0.00 & 257.15 & 0.00 & 54.02 & 43.83 & 3.34 & 77.69 & 4.92 & 6.15 & 712.47 & 6.89 & 0.00 & 1.00 \\ 
  7826 & 5.00 & 2.62 & 7.62 & 2.19 & 0.00 & 366.08 & 0.01 & 20.04 & 40.75 & 2.80 & 61.34 & 2.68 & 4.17 & 606.61 & 4.58 & 0.00 & 1.00 \\ 
  7857 & 18.90 & 4.56 & 65.33 & 2.04 & 8.56 & 518.68 & 0.07 & 0.00 & 50.38 & 8.03 & 2.60 & 2.92 & 2.90 & 689.37 & 3.46 & 0.00 & 0.00 \\ 
  7923 & 19.80 & 4.38 & 46.36 & 1.30 & 0.00 & 394.07 & 1.95 & 28.72 & 58.44 & 5.83 & 34.03 & 4.37 & 4.48 & 713.97 & 5.44 & 0.00 & 1.00 \\ 
  7951 & 20.00 & 46.68 & 25.75 & 1.68 & 8.64 & 110.43 & 0.78 & 83.50 & 54.55 & 19.15 & 36.29 & 8.67 & 4.01 & 1037.07 & 8.17 & 0.00 & 0.00 \\ 
  7995 & 2.70 & 63.84 & 61.68 & 1.15 & 16.23 & 34.77 & 0.75 & 33.82 & 49.72 & 25.12 & 8.35 & 6.58 & 3.16 & 1247.05 & 9.39 & 0.00 & 0.00 \\ 
  8034 & 1.77 & 4.08 & 7.36 & 1.13 & 0.00 & 276.15 & 0.10 & 104.89 & 58.44 & 5.83 & 34.03 & 4.37 & 4.48 & 713.97 & 5.44 & 0.00 & 1.00 \\ 
  8037 & 74.00 & 0.99 & 67.05 & 3.81 & 0.00 & 361.99 & 0.03 & 23.08 & 58.44 & 5.83 & 34.03 & 4.37 & 4.48 & 713.97 & 5.44 & 0.00 & 1.00 \\ 
  8058 & 1.59 & 8.52 & 6.24 & 1.20 & 0.00 & 273.43 & 0.28 & 189.09 & 58.44 & 5.83 & 34.03 & 4.37 & 4.48 & 713.97 & 5.44 & 0.00 & 1.00 \\ 
  8794 & 7.00 & 21.35 & 35.70 & 1.54 & 7.37 & 400.45 & 1.85 & 23.71 & 40.75 & 2.80 & 61.34 & 2.68 & 4.17 & 606.61 & 4.58 & 0.00 & 0.00 \\ 
  8883 & 6.40 & 9.01 & 76.36 & 1.51 & 0.03 & 535.46 & 0.00 & 68.02 & 51.67 & 23.34 & 13.99 & 6.28 & 3.12 & 1101.03 & 6.15 & 0.00 & 1.00 \\ 
  9068 & 14.60 & 4.25 & 87.91 & 1.26 & 0.00 & 336.82 & 0.17 & 94.64 & 47.92 & 24.18 & 6.59 & 5.88 & 2.87 & 1205.86 & 6.80 & 0.00 & 1.00 \\ 
  9078 & 6.10 & 4.68 & 59.66 & 1.83 & 4.72 & 399.28 & 3.90 & 0.00 & 58.44 & 5.83 & 34.03 & 4.37 & 4.48 & 713.97 & 5.44 & 0.00 & 0.00 \\ 
  9191 & 24.10 & 3.49 & 79.27 & 1.30 & 0.00 & 405.93 & 0.64 & 74.12 & 58.44 & 5.83 & 34.03 & 4.37 & 4.48 & 713.97 & 5.44 & 0.00 & 1.00 \\ 
  9320 & 2.91 & 8.02 & 49.26 & 2.12 & 4.61 & 309.87 & 0.04 & 74.69 & 47.95 & 8.32 & 29.33 & 5.89 & 3.94 & 833.54 & 9.01 & 1.00 & 0.00 \\ 
  9443 & 1.50 & 4.68 & 54.18 & 1.76 & 2.29 & 195.83 & 2.24 & 24.15 & 54.55 & 19.15 & 36.29 & 8.67 & 4.01 & 1037.07 & 8.17 & 1.00 & 0.00 \\ 
  9617 & 68.58 & 0.28 & 13.11 & 1.09 & 0.00 & 399.40 & 0.00 & 1.85 & 58.44 & 5.83 & 34.03 & 4.37 & 4.48 & 713.97 & 5.44 & 0.00 & 1.00 \\ 
  9797 & 8.50 & 11.69 & 15.68 & 1.48 & 7.98 & 385.99 & 0.37 & 9.53 & 40.75 & 2.80 & 61.34 & 2.68 & 4.17 & 606.61 & 4.58 & 1.00 & 0.00 \\ 
  9885 & 13.40 & 2.44 & 66.64 & 2.25 & 5.09 & 410.52 & 0.24 & 45.28 & 58.44 & 5.83 & 34.03 & 4.37 & 4.48 & 713.97 & 5.44 & 0.00 & 0.00 \\ 
  10231 & 25.91 & 3.42 & 72.33 & 1.96 & 4.70 & 385.27 & 0.69 & 0.00 & 50.38 & 8.03 & 2.60 & 2.92 & 2.90 & 689.37 & 3.46 & 0.00 & 0.00 \\ 
   \hline
\end{tabular}
}
\end{table}